\documentclass[12pt]{iopart}

\newcommand{\nls}[1]{\chi_{\scriptscriptstyle B}^{(#1)}}

\include{graphicx}
\begin{document}

\title{Exploring the QCD landscape with high-energy nuclear collisions}

\author{Bedangadas Mohanty}

\address{Variable Energy Cyclotron Centre, 1/AF, Bidhan Nagar, Kolkata - 700064, India}
\ead{bmohanty@vecc.gov.in}
\begin{abstract}
Quantum chromodynamics (QCD) phase diagram 
is usually plotted as temperature ($T$) versus the chemical potential 
associated with the conserved baryon number ($\mu_{B}$). Two fundamental properties 
of QCD, related to confinement and chiral symmetry, allows for two corresponding 
phase transitions when $T$ and $\mu_{B}$ are varied. Theoretically the phase diagram 
is explored through non-perturbative QCD calculations on lattice.  The energy scale 
for the phase diagram ($\Lambda_{\mathrm {QCD}}$ $\sim$ 200 MeV) is such that 
it can be explored experimentally by colliding nuclei at varying beam energies in 
the laboratory. In this paper we review some aspects of the QCD phase structure 
as explored through the experimental studies using high energy nuclear collisions. 
Specifically, we discuss three observations related to the formation of a strongly 
coupled plasma of quarks and gluons in the collisions, experimental search for 
the QCD critical point on the phase diagram and freeze-out properties of the hadronic phase. 
\end{abstract}

\vspace{2pc}

\section{QCD Phase Diagram}
\begin{figure}
\begin{center}
\includegraphics[scale=0.7]{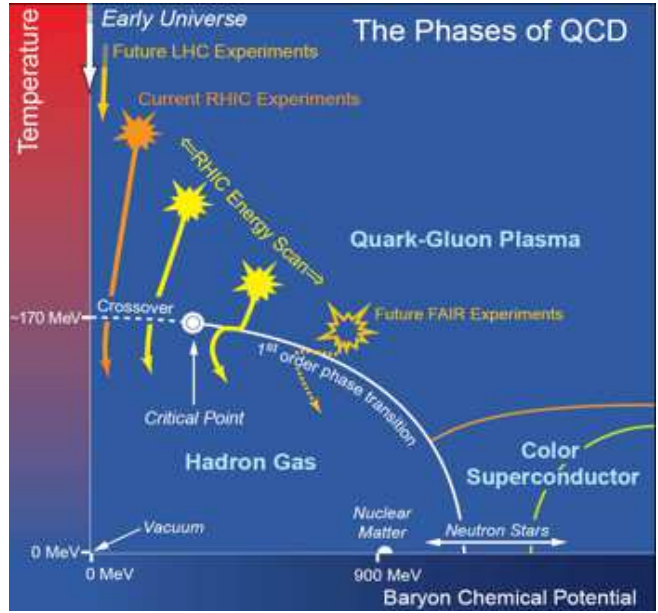}
\caption{(Color online) Typical phase diagram of QCD. See text for details. Figure taken from Ref.~\cite{usnsac}.}
\label{fig1}
\end{center}
\end{figure}

Physical systems undergo phase transitions when external parameters 
such as the temperature ($T$) or a chemical potential ($\mu$) are changed. 
A phase diagram provides intrinsic knowledge on the structure of the matter 
under study. In other words, it tells us how matter organizes itself 
under external conditions at a given degrees of freedom. The theory of 
strong interactions, QCD, predicts that nuclear matter at high temperature 
and/or density makes a transition from a state where quarks and gluons are 
confined and chiral symmetry is broken to a state where quarks and gluons 
are de-confined and chiral symmetry is restored~\cite{firstpapers}. QCD has several conserved 
quantities: baryon number ($B$), electric charge ($Q$), and strangeness ($S$). 
Each of these is associated with a chemical potential. As a result, the QCD 
phase diagram  is four-dimensional.  $\mu_{\rm Q}$ and $\mu_{\rm S}$ are 
relatively small compared to $\mu_{\rm B}$ (baryonic chemical potential) 
in high energy nuclear collisions~\cite{starpidprc}. The $T$ and $\mu_{\rm B}$ 
are varied in a typical QCD phase diagram as shown in Fig.~\ref{fig1}~\cite{usnsac}. 
At high temperature and density the phase is governed 
by quark and gluon degrees of freedom and is commonly referred to as the 
Quark Gluon Plasma (QGP)~\cite{shuryakphysreport}. At large densities and low temperatures 
other interesting phases related to neutron star~\cite{neutronstarreview} and 
color superconductivity~\cite{colorsupercond} starts to appear.
These QCD transitions which occurred in the early universe have the right energy scale 
to be accessible by the experiments. The  Fig.~\ref{fig1} shows the parts of the
phase diagram explored by several accelerator based experimental programs. 
Experimentally, this is done by varying the beam energy. Both, $T$ and $\mu_{\mathrm B}$,  
vary as the function of the center-of-mass energy ($\sqrt{s_{\rm {NN}}}$)~\cite{peternature}. 
This strategy  is being followed by experimental programs at 
Large Hadron Collider (LHC) at CERN, 
Relativistic Heavy Ion Collider (RHIC) at BNL and Super Proton Synchrotron (SPS) at CERN, and
will be followed at Facility for Antiproton and Ion Research (FAIR) at GSI and Nuclotron-based Ion Collider 
fAcility (NICA) at JINR.  Among these experiments, we discuss in section 2, selected results 
from RHIC which provides evidence for the formation of a QGP~\cite{rhicwhitepapers}. 

Theoretically, finite temperature lattice QCD calculations at 
$\mu_{\rm B}$ = 0 suggest a cross-over~\cite{latticenature} above a 
temperature, $T_{\rm c}$, of about  170 to 190 MeV from a system 
with hadronic degrees of freedom to a QGP~\cite{tc}. 
At large $\mu_{\rm B}$, several QCD based calculations find the 
quark-hadron phase transition to be of first order~\cite{firstorder}. 
Going towards the smaller $\mu_{\rm B}$ region, the point 
in the QCD phase plane ($T$ vs. $\mu_{\rm B}$) where the first order 
phase transition ends is the QCD Critical Point (CP)~\cite{criticalpointtheory}. 
The focus in the coming decade would be on attempts to locate the CP 
both experimentally and theoretically. Current theoretical calculations 
are highly uncertain about the location of the CP. This is primarily because 
the lattice QCD calculations at finite $\mu_{\rm B}$ face numerical challenges. 
The experimental plan (discussed in section 3) is to vary the 
$\sqrt{s_{\rm {NN}}}$ of heavy-ion 
collisions to scan the phase plane and, at each energy, search for 
signatures of the CP that might survive the evolution of the system. 
In the last section of the review we discuss the thermodynamic properties of the 
hadronic phase. Finally we end with a summary of our current understanding 
of phase diagram and a brief outlook.

\section{Establishing the Partonic Phase at RHIC} 

\subsection{Strangeness enhancement and formation of a gluon rich plasma}

Enhancement of strange hadron production in high energy heavy-ion collisions~\cite{seqgp} 
due to formation of QGP is one of the four classic signatures in 
this field. The other three being, enhanced direct photon~\cite{directphoton} and dilepton~\cite{dilepton} production and 
suppression of J/$\Psi$ production~\cite{jpsisatz} in heavy-ion collisions relative to $p$+$p$ collisions. In a QGP, 
thermal $s$ and $\bar{s}$ quarks can be produced by  gluon-gluon interactions~\cite{shor}. These 
interactions could occur very rapidly and the $s$-quark abundance would equilibrate. During 
hadronisation, the $s$ and $\bar{s}$ quarks from the plasma coalesce to form $\phi$ mesons. 
Production  by this process is not suppressed as per the OZI (Okubo-Zweig-Izuka) rule~\cite{ozi}. 
This, coupled with large abundances of strange quarks in the plasma, may lead to a dramatic increase 
in the production of $\phi$ mesons and other strange hadrons relative to non-QGP $p$+$p$ collisions.

\begin{figure}
\begin{center}
\includegraphics[scale=0.5]{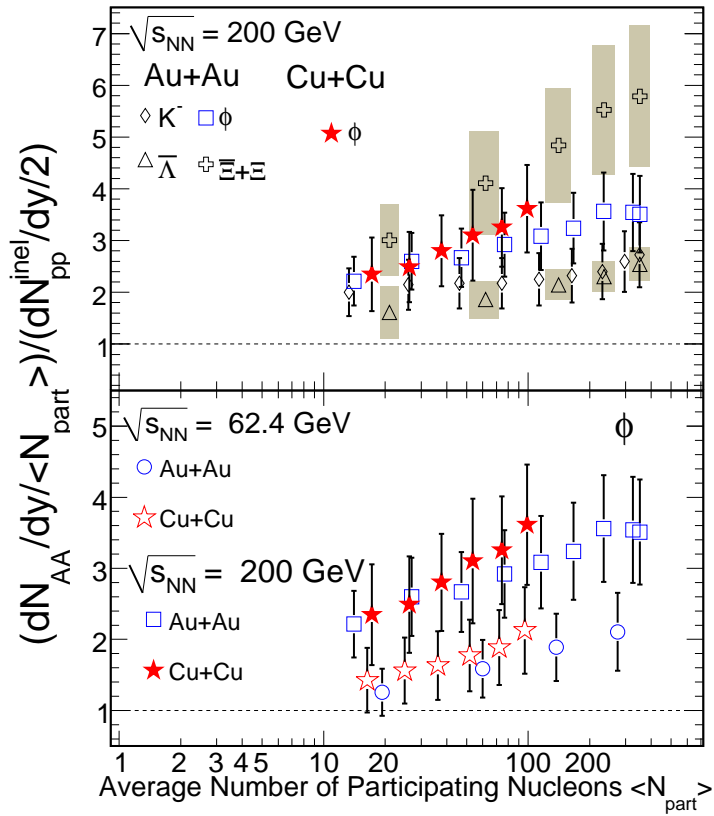}
\caption{(Color online) Upper panel: The ratio of the yields of 
$K^{-}$, $\phi$, $\bar{\Lambda}$ and  $\Xi+\bar{\Xi}$
normalized to average number of participating nucleons ($\langle N_{\mathrm {part}} \rangle$) 
in nucleus-nucleus collisions to corresponding yields in inelastic $p$+$p$ 
collisions as a function of $\langle N_{\mathrm {part}} \rangle$ at 200 GeV.
Lower panel: Same as above for $\phi$ mesons in Cu+Cu collisions at 200 and 62.4 GeV.
The $p$+$p$ collision data at 200 GeV are from STAR experiment and 
at 62.4 GeV from ISR~\cite{isr}. The error bars shown represent the statistical 
and systematic errors added in quadrature. Figure is taken from Ref.~\cite{phiplb}.
}
\label{phi}
\end{center}
\end{figure}
Such predictions of relative enhancement of strange hadron was challenged by an alternate idea 
of canonical suppression of strangeness  in small systems being the source of strangeness 
enhancement in $\Lambda$, $\Xi$ and $\Omega$ hadrons in high energy heavy-ion collisions~\cite{canonical}. 
The strangeness conservation laws require 
the production of an $\bar{s}$-quark for each $s$-quark in the strong interactions. The main argument 
in such canonical models is that the energy and space time extensions in smaller systems may not be 
sufficiently large. This leads to a suppression of open-strange hadron production in small collision systems.
These statistical models fit the data reasonably well~\cite{canonical_expt}. These models predict
two things: (a) strangeness enhancement in nucleus-nucleus collisions, relative to $p$+$p$ collisions, 
should increase with the valence strange quark content of the hadrons and (b) the enhancement is predicted to 
decrease with increasing beam energy~\cite{canonical_prediction}.  
Discriminating between the two scenarios, QGP versus the canonical suppression, 
using the experimental data on $K^{\pm}$, $\Lambda$, $\Xi$ and $\Omega$ hadrons has been, 
to some extent, ambiguous. Enhancement of $\phi (s\bar{s})$ production (zero net-strangeness, hence
not subjected to the canonical suppression) in heavy-ion collisions relative to $p$+$p$ collisions 
would clearly indicate the formation of a gluon rich QGP in these collisions. This would then rule 
out canonical suppression scenario. 

Figure~\ref{phi} shows the enhancement of strange hadron production at RHIC~\cite{phiplb}. 
Upper panel shows the ratio of strange hadron production normalized 
to $\langle N_{\mathrm {part}} \rangle$ in nucleus-nucleus collisions relative to corresponding
results from $p$+$p$ collisions at 200 GeV. The results are plotted as a function 
of $\langle N_{\mathrm {part}} \rangle$. $K^{-}$, $\bar{\Lambda}$ and  $\Xi+\bar{\Xi}$ are 
found to exhibit an enhancement (value $>$ 1) that increases with the number of strange valence quarks.
Furthermore, the observed enhancement in these open-strange hadrons increases with collision centrality, 
reaching a maximum for the most central collisions. However, the enhancement of $\phi$ meson production shows a
deviation from the ordering in terms of the number of strange constituent quarks. 
More explicitly, this enhancement is larger than for $K^{-}$ and $\bar{\Lambda}$, at the
same time being smaller than in case of $\Xi+\bar{\Xi}$.
Despite being different particle types (meson-baryon) and having different masses,
the results for $K^{-}$ and $\bar{\Lambda}$ are very similar in the entire centrality region studied. 
This rules out a baryon-meson effect as being the reason for the deviation of $\phi$ mesons 
from the number of strange quark ordering seen in Fig.~\ref{phi} (upper panel).
The observed deviation is also not a mass effect as the enhancement in $\phi$ meson production 
is larger than that in $\bar{\Lambda}$ (which has mass close to that of the $\phi$).
Further in heavy-ion collisions, the production of $\phi$ mesons is not canonically suppressed 
due to its $s$$\bar{s}$ structure. 

The observed enhancement of $\phi$ meson production then is a clear indication for the formation
of a dense partonic medium being responsible for the strangeness enhancement in Au+Au 
collisions at 200 GeV. The observed enhancement in $\phi$ meson production being related to medium density 
is further supported by the energy dependence shown in the lower panel of  Fig.~\ref{phi}. 
The $\phi$ meson production relative to $p$+$p$ collisions is larger at higher beam energy, 
a trend opposite to that predicted in canonical models for other strange hadrons.  
In addition measurements have shown that $\phi$ meson production is not from the coalescence of 
$K\bar{K}$ and is minimally affected by re-scattering effects in the medium~\cite{phisqm}. 
Measurements also indicate that 
$\phi$ mesons are formed from the coalescence of seemingly thermalized strange quarks~\cite{phiprl}.
All these observations put together indicate the formation of a dense partonic medium in 
heavy-ion collisions where strange quark production is enhanced. 
This in turn suggests that the observed centrality 
dependence of the enhancement for other strange hadrons (seen in Fig.~\ref{phi}) is likely to be related to
the same reasons as in the case of the $\phi$ meson, that it is due to the formation of a dense gluon rich 
partonic medium in the collisions. These experimental data rule out the possibility of 
canonical suppression being the only source of the observed strangeness enhancement 
at $\sqrt{s_{NN}}$= 200 GeV.

\subsection{Jet Quenching and highly opaque medium}

One of the most exciting results to date at RHIC is the discovery
of suppression in the production of high transverse momentum ($p_{T}$) mesons in nucleus-nucleus
collisions compared to corresponding data from the binary collision scaled p+p collisions~\cite{jetquenchingexpt}. 
This has been interpreted in terms of energy loss of partons in QGP.
This phenomena is called as the jet quenching in a dense partonic matter~\cite{jetquenching}. 
The energy loss by energetic partons traversing the dense medium formed in
high-energy heavy-ion collisions is predicted to be proportional to both the initial
gluon density~\cite{vitev_density} and the lifetime of the dense matter~\cite{xnwang_lifetime}. 
The results on high-$p_{T}$ suppression are usually presented in terms of the nuclear modification
factor ($R_{AA}$), defined as:

\begin{equation}
R_{AA}=\frac{dN_{AA}/d\eta d^2 p_{T}} {T_{AB} d\sigma_{NN}/d\eta d^2 p_{T}}
\label{eq:RAA}
\end{equation}
where the overlap integral $T_{AB} = N_{binary}/\sigma_{inelastic}^{pp}$ with $N_{binary}$ being the
number of binary collisions commonly estimated from Glauber model calculation~\cite{glauber}.

\begin{figure}
\begin{center}
\includegraphics[scale=0.7]{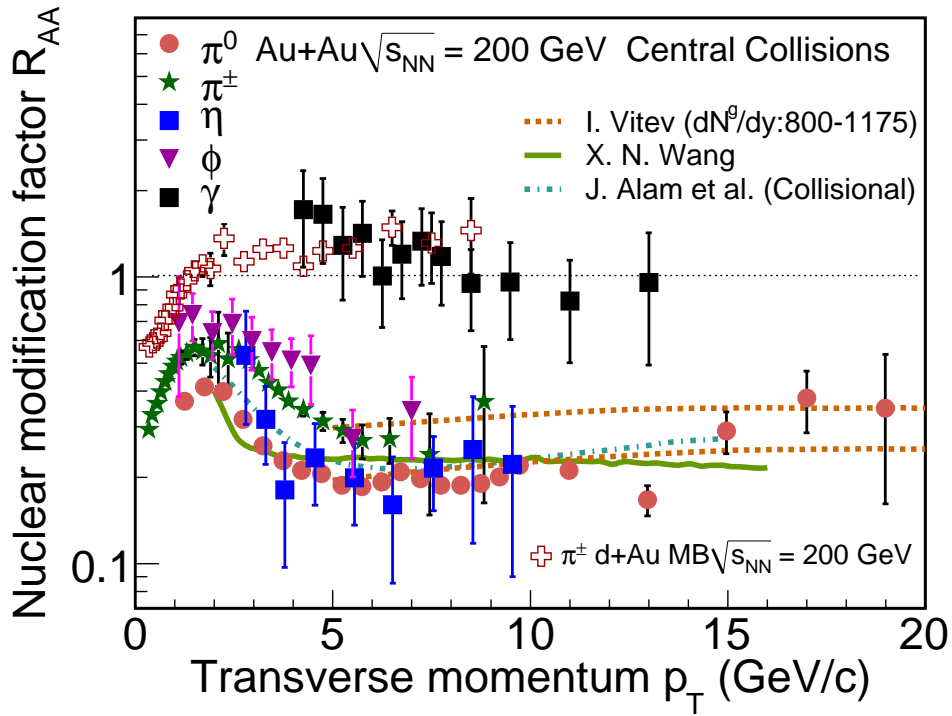}
\caption{ (Color online) Compilation of the nuclear modification factor ($R_{\mathrm {AA}}$) for mesons and 
direct photons as measured in RHIC experiments at midrapidity for central Au+Au
collisions at $\sqrt{s_{\mathrm {NN}}}$ = 200 GeV. Also shown are the $R_{\mathrm {dAu}}$ for
charged pions for $\sqrt{s_{\mathrm {NN}}}$ = 200 GeV. 
The lines are results from various model calculations. See text for more details.}
\label{raa}
\end{center}
\end{figure}
In Fig.~\ref{raa} we show the RHIC data for the $R_{AA}(p_{\mathrm T})$,
for various produced mesons~\cite{exptraa} and direct photons~\cite{photonraa} in central Au+Au collisions 
at midrapidity. A large suppression in high  $p_{\mathrm T}$ meson production is observed, and those for 
$\pi^0$'s being almost flat at $R_{AA} \simeq 0.2$ up to 20 GeV/$c$.
The figure also shows that the level of suppression for $\pi^0$'s, $\eta$'s and $\phi$-mesons are 
very similar, which supports the conclusion that the suppression occurs in the partonic phase, 
not in the hadronic phase. This strong suppression of meson production is in contrast to the behavior of
direct photons, also shown in the figure. The direct photons follow
binary scaling (i.e. $R_{AA} \simeq 1$) or no suppression. This is a strong evidence that the
suppression is not an initial state effect, but a final state effect caused
by the high density medium with color charges created in the collision. This is further
consolidated by a demonstration through a controlled experiment using deuteron
on Au ion collisions, which gave a $R_{dAu}(p_{\mathrm T})$ $\sim$ 1 for $\pi^{\pm}$
at midrapidity and high $p_{\mathrm T}$~\cite{rdau}.

The various curves in the Fig.~\ref{raa} represents different model calculations.
The dashed curve shows a theoretical prediction using the GLV parton energy loss model \cite{vitev_density} . 
The model assumes an initial parton density $dN/dy=800-1100$, which corresponds to an energy
density of approximately 5-15 GeV/fm$^3$. The lower dashed curves are for higher gluon density. 
The precision high $p_{\mathrm T}$  data at RHIC has been used to characterize medium density fairly accurately. 
The conclusion being that the medium formed in central Au+Au collisions at RHIC
has a high degree of opacity~\cite{phenixopacity}.
In addition, theoretical studies suggest that for a given  initial density, 
the $R_{AA}$($p_{\mathrm T}$) values are also sensitive to the lifetime ($\tau$) of dense matter 
formed in heavy-ion collisions~\cite{xnwang_lifetime}.  The solid curves are predictions from
Ref.~\cite{xnwang_lifetime} at  $\sqrt{s_{\mathrm {NN}}}$ = 200 GeV with $\tau$ = 10 fm/$c$ 
(i.e. larger than the typical system size of $\sim$ 6-7 fm). The parton energy loss calculations
discussed above attributes the opacity to plasma induced radiation of gluons, much like ordinary
bremsstrahlung of photons by electrons. However, the quantitatively large suppression pattern observed 
at high $p_{\mathrm T}$, for both light hadrons and those involving heavy quarks~\cite{nonphotonic}, 
showed that the mechanism of energy loss is far from being a settled issue, namely, the relative
contribution of radiative and collisional forms. As an example, shown in Fig.~\ref{raa} is 
a comparison of the data to theoretical results (dot-dash curves) 
on $R_{AA}$ from models that consider only collisional energy loss~\cite{alam}. 
This model gives $R_{AA}$ values at high $p_{\mathrm T}$ close to the measured values 
and similar to corresponding values from models having only a radiative mechanism for 
parton energy loss.

Leading particle measurements, such as the ones shown in Fig.~\ref{raa} suffer from a number of limitations. 
(a) Leading hadrons come from a mixture of parent quarks and gluons. (b) As a fragmentation product, 
the energy of a leading hadron is not a perfect proxy for the energy of the parent parton as
it samples a wide range of partonic energies. In future we should look forward to potentially three interesting
measurements. (i) The $\gamma$-jet process potentially provides access to the underlying scattered parton's 
energy. Measurements of the distribution of particles from the jet opposite, in azimuth, to the tagged photon reveals how 
much energy was lost, and how it was redistributed, by the colored parton as it traversed the medium~\cite{tagphoton}.
(ii) Another method is through the full reconstruction of jets in heavy-ion collisions~\cite{jethi}. Beyond producing 
a far better proxy for the energy of the parent
parton than a leading hadron, this technique allows 
one to trace the evolution of energy flow in
directions both longitudinal and transverse to the 
direction of the parent parton. Both these methods are under active pursual at RHIC.
(iii) Another important feature of jet quenching is provided by partonic identity. While it is difficult
to disentangle light quarks from gluons, especially in a heavy-ion environment, charm and bottom
can be easily tagged by the existence of a charmed or bottom hadron in the final state. Due to
their large masses the charm and bottom quarks are predominantly produced via hard scattering in
the initial stage of the high-energy heavy-ion collision. The final state spectra can therefore serve
as a sensitive tool to probe in-medium rescattering and interactions responsible for thermalization.
This will also allow to study the non-Abelian feature of QCD that results in the gluons losing more
energy than quarks in the medium~\cite{colorfactor}.
Plans are in place to measure the cross-sections and transverse momentum spectra of hadrons with open
and hidden heavy flavor at RHIC with new detector upgrades. This will also provide useful data to
understand the different mechanisms of energy loss: collisional versus radiative.

\subsection{Partonic collectivity and low viscosity}

Elliptic flow, $v_{2}$, is an observable which is thought to reflect
the conditions from the early stage of the collisions \cite{v2Early}. 
In non-central heavy-ion collisions, the initial spatial anisotropy of the 
overlap region of the colliding nuclei is transformed into an anisotropy 
in momentum space through interactions between the particles. As the system expands 
it becomes more spherical, thus the driving force quenches itself. 
Therefore the elliptic flow is sensitive to the collision dynamics 
in the early stages. It is measured, by calculating $\langle\cos(2(\phi-\Psi))\rangle$, 
where $\phi$ is the azimuthal angle of the produced particles and $\Psi$ is the azimuthal 
angle of the impact parameter, and angular brackets denote an average over many particles and events.
\begin{figure}
\begin{center}
\includegraphics[scale=0.35]{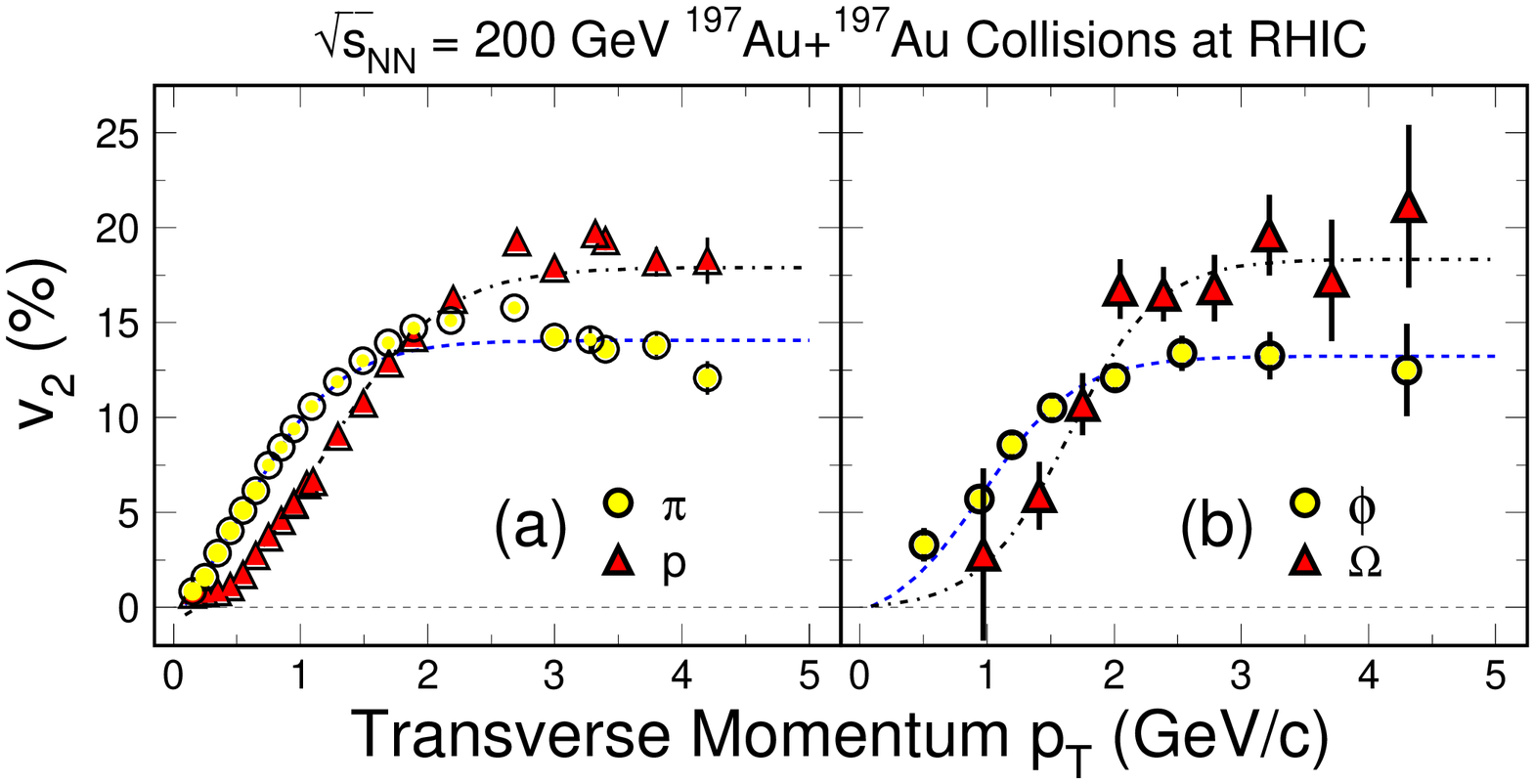}
\includegraphics[scale=0.35]{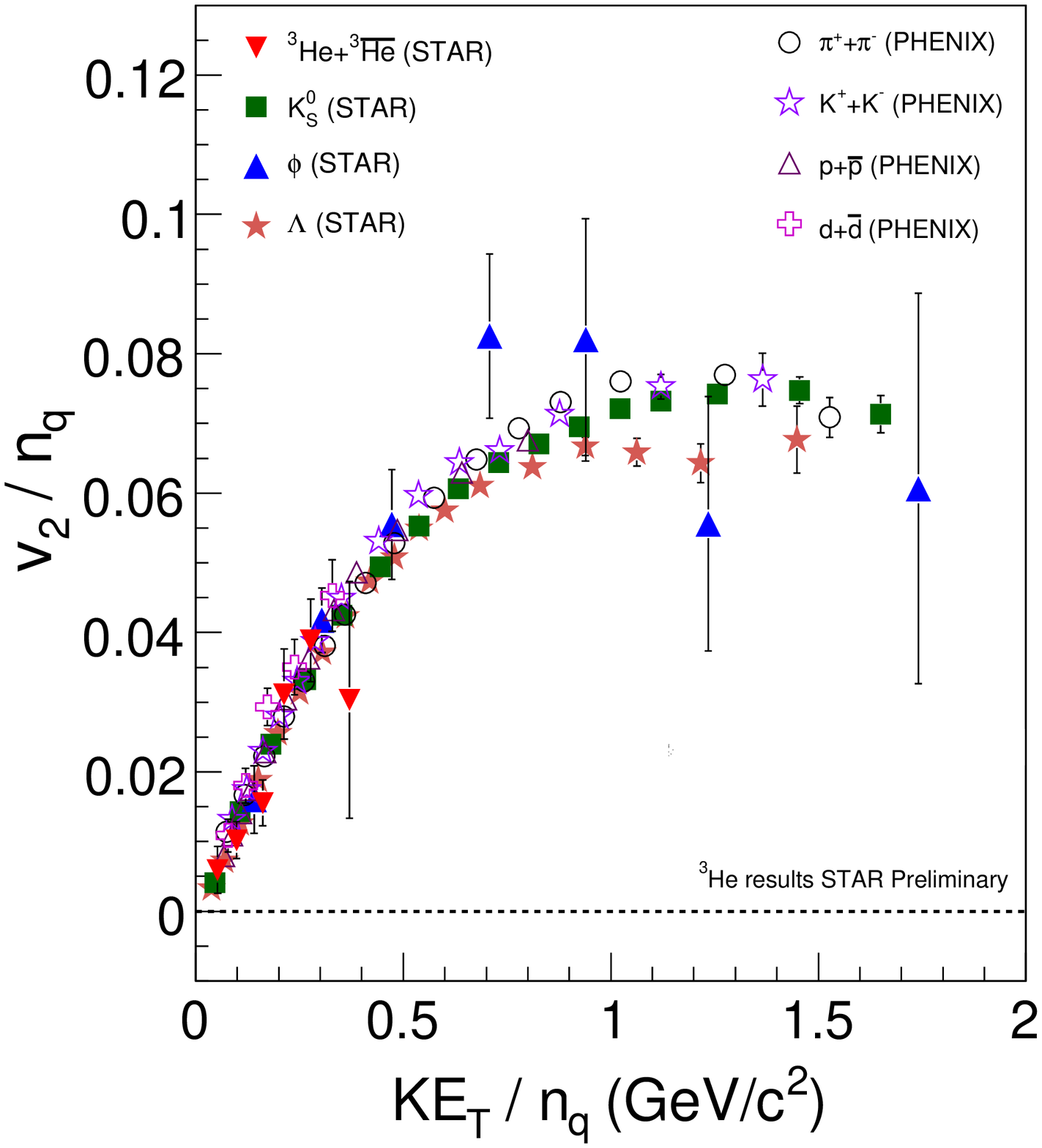}
\includegraphics[scale=0.45]{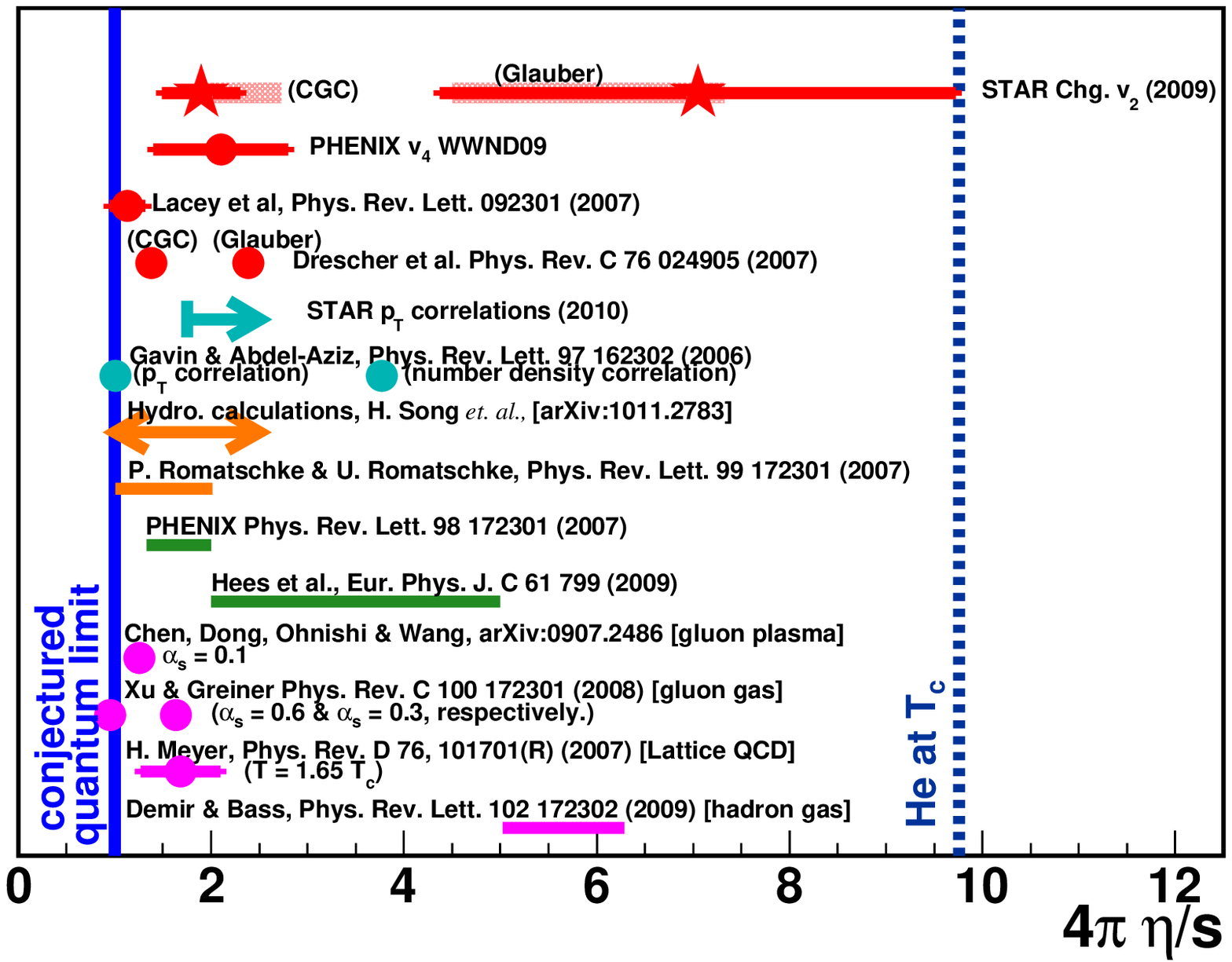}
\caption{(Color online) Top left plot: The elliptic flow
$v_2$ for (a) light quark hadrons and (b) strange quark hadrons. 
The data is from the minimum bias Au+Au collisions at midrapidity for $\sqrt{s_{\rm {NN}}}$ = 200 GeV~\cite{starflowqm09}. 
Top right plot: Compilation of the number of constituent quark scaled $v_2$ as a function 
of the scaled transverse kinetic energy~\cite{chita}. Bottom plot: Compilation of
$\eta/s$ extracted from various measurements in heavy-ion collisions at RHIC. }
\end{center}
\label{v2}
\end{figure}

Figure~4 top left panel shows the RHIC results on $v_2$ of light quark ($\pi$ and $p$)
and strange quark ($\phi$ and $\Omega$) carrying hadrons in Au+Au collisions at 200 GeV~\cite{starflowqm09}. 
Three very distinct experimental observations can be made. (a) At low $p_{\mathrm T}$
($<$ 2 GeV/$c$) the heavier hadrons have smaller $v_{2}$. Such a mass ordering is
expected in hydrodynamics calculations of $v_{2}$($p_{\mathrm T}$) for  identified 
particles \cite{Hydro}. (b) At the intermediate $p_{\mathrm T}$ range
of 2-5 GeV/$c$ it is observed that baryons have higher $v_{2}$ than mesons. The $\phi$-meson
$v_{2}$ plays a crucial role in establishing this baryon-meson difference. Such a 
separation of baryons and mesons in the intermediate $p_{T}$ range has been also observed 
in measurements of the nuclear modification factor, $R_{CP}$~\cite{RCP}. These results are
consistent with calculations from quark recombination models~\cite{reco} implying the
de-confinement of the system prior to hadronisation. (c) Comparison between the $v_{2}$
results for light quark carrying hadrons to those from strange quark carrying hadrons
indicates both types of hadrons show similar 
magnitude of $v_{2}$. The multi-strange hadrons ($\phi$ and $\Omega$) have
relatively low hadronic interaction cross sections and freeze-out early, hence are considered as the 
most promising probes of the early stages of the collision. All the above results
indicates that a substantial amount of collectivity has been developed at the 
partonic stage of the heavy-ion collisions. In fact none of the available hadronic models
are able to account for the observed magnitude of $v_{2}$ at $\sqrt{s_{\mathrm {NN}}}$ = 200 GeV. 
Only models which have additional partonic interactions explain the $\langle v_{2} \rangle$ values~\cite{nasim}.

When elliptic flow $v_2$ is plotted versus transverse kinetic 
energy ($m_{\rm T} - m_0$), both divided by the number of constituent 
quarks, the $v_2$ for all identified hadrons as well as light nuclei below 
($m_{\rm T} - m_0$) $\sim$ 1 GeV/$c^2$ falls on a universal curve~\cite{ncqexpt}. 
$m_{0}$ is the mass of the particle.
This scaling behavior as shown in Fig.4 further strengthens the 
evidence for formation of partonic matter during the 
Au + Au collision process at 200 GeV. It is very hard to explain this 
observed pattern in a scenario where only hadronic matter exists 
throughout the interaction, whereas the hypothesis of coalescence 
of hadrons from de-confined quarks offers a ready explanation. 
Turn-off of the scaling at a given beam energy would indicate the hadronic 
side of the phase boundary. 

The $v_2$ measurements of light quark carrying hadrons, the nuclear
modification factor and $v_2$ for heavy quark carrying hadrons and the
differential $p_{\mathrm T}$ correlations for charged hadrons have been 
used to extract information of a dimensionless ratio, shear viscosity to
entropy ($\eta/s$), for the medium formed in heavy-ion collisions at RHIC~\cite{etabys}.
Figure~4 bottom panel shows the compilation of $\eta/s$ extracted
from various measurements~\cite{aihong}. It is observed to lie between that conjectured
from quantum theory ($\eta/s$ $\sim$ 1/4$\pi$) and those for liquid Helium at $T_{c}$.
Such a low value of $\eta/s$ (within a factor 1-10 of the quantum limit) 
indicates that the matter formed in heavy-ion collisions at RHIC has low
viscosity, hence is a strongly coupled system. 

\section{QCD critical point and thermalization}

The CP is a landmark point in the QCD phase diagram, observation of which
will make the QCD phase diagram a reality. A close collaboration between the experiments and theory 
perhaps will lead to its discovery.
The first step in this process is to establish an observable for CP which can be measured experimentally 
and can be related to QCD calculations. In this context, it is important 
to recall that for a static, infinite medium, the correlation 
length ($\xi$) diverges at the CP. $\xi$ is related to various moments of the distributions of conserved 
quantities such as net-baryons, net-charge, and net-strangeness~\cite{volker}. Typically variances 
($\sigma^2$ $\equiv$ $\left\langle (\Delta N)^2 \right\rangle$; $\Delta N = N - M$; $M$ is the mean) 
of these distributions are related to $\xi$ as  $\sigma^2$ $\sim$ $\xi^2$~\cite{stephanovprd}. 
Finite size and time effects in heavy-ion collisions put constraints on the 
values of $\xi$. A theoretical calculation suggests $\xi$ $\approx$ 2-3 fm 
for heavy-ion collisions~\cite{krishnaxi}. It was recently shown that higher 
moments of distributions of conserved quantities, measuring deviations from a 
Gaussian, have a sensitivity to CP fluctuations that is better than that of $\sigma^2$, 
due to a stronger dependence on $\xi$~\cite{stephanovmom}. 
The numerators in skewness ($\it {S}$ = $\left\langle (\Delta N)^3 \right\rangle/\sigma^{3}$) goes 
as $\xi^{4.5}$ and  kurtosis ($\kappa$ = [$\left\langle (\Delta N)^4 \right\rangle/\sigma^{4}$] - 3) 
goes as $\xi^7$. A crossing of the phase boundary can manifest itself by 
a change of sign of $\it{S}$ as a function of energy density~\cite{stephanovmom,asakawa}.

\begin{figure}[htp]
\begin{center}
\includegraphics[scale=0.45]{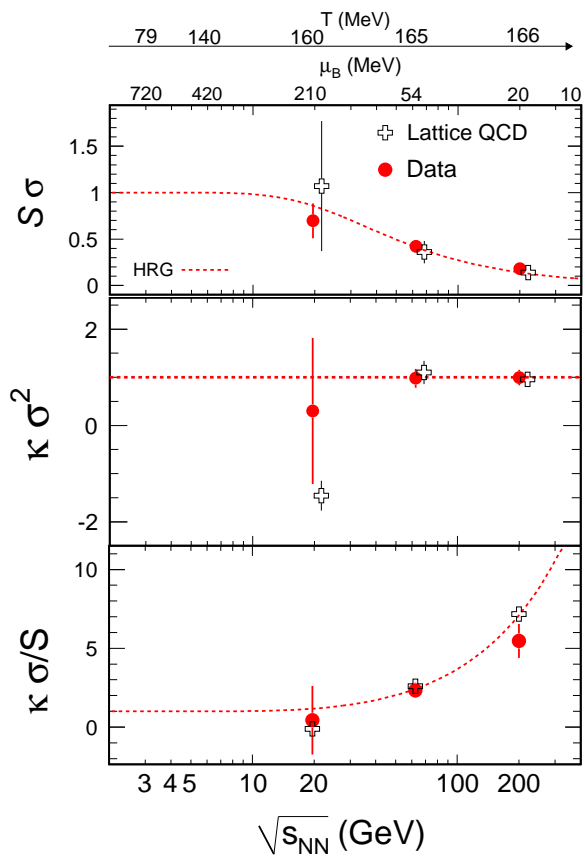}
\caption{(Color online) $\sqrt{s_{\mathrm {NN}}}$ dependence of $S\sigma$, $\kappa$$\sigma^2$ 
and $\frac{\kappa\sigma}S$ for
net-proton distributions measured at RHIC~\cite{starkurtosisprl}. The results are compared Lattice QCD
calculations~\cite{latticesus} and HRG model~\cite{hrg}. Also shown on the top are the temperature and baryon chemical
potential values at chemical freeze-out extracted from particle yields using a thermal model.
}
\label{qcp}
\end{center}
\end{figure}
The first connections between QCD calculations and experiment has been recently made~\cite{starkurtosisprl}.
Lattice calculations and QCD-based models show that moments of net-baryon distributions are related to baryon 
number ($\Delta N_{\mathrm B}$) susceptibilities ($\chi_{\mathrm B} = \frac{\left\langle (\Delta N_{\mathrm B})^{2}\right\rangle}{VT}$; {\it V} is the volume)~\cite{latticesus,latticesus1}. Then one can construct ratios such as: 
\begin{eqnarray}
\nonumber
  S\sigma = \frac{T\nls3}{\nls2}, \kappa\sigma^2= \frac{T^2\nls4}{\nls2} ~~{\rm and}~~ \frac{\kappa\sigma}S = \frac{T\nls4}{\nls3},
\label{ratios}\end{eqnarray}
which do not contain the volume and therefore provide a direct and convenient
comparison of experiment and theory. In the above expressions the left hand side of 
each equality can be measured in an experiment whereas the right hand side can
be calculated by lattice QCD. Close to the CP models predict the $\Delta N_{\mathrm B}$ 
distributions to be non-Gaussian and susceptibilities to diverge causing the 
experimental observables to have large values. The experimental values should also be 
compared to those expected from statistics, for example if the $p$ and $\bar{p}$ 
distributions are individually Poissonian then $\kappa$$\sigma^2$ for net-protons is unity. 

Experimentally measuring event-by-event net-baryon number is difficult. However, the net-proton multiplicity 
($N_{\mathrm p} - N_{\bar{\mathrm p}}$ = $\Delta N_{\mathrm p}$) distribution is measurable. 
Theoretical calculations have shown  that $\Delta N_{\mathrm p}$ fluctuations reflect 
the singularity of the charge and baryon number susceptibility as expected at 
the CP~\cite{hatta}. Non-CP model calculations show that the inclusion of other baryons does not add 
to the sensitivity of the observable~\cite{starkurtosisprl}.

Figure~\ref{qcp} shows the energy dependence of $S\sigma$, $\kappa$$\sigma^2$ 
and $\frac{\kappa\sigma}S$ for $\Delta N_{\mathrm p}$, 
compared to lattice QCD~\cite{latticesus} and Hadron Resonance Gas (HRG)  model which does not include a CP~\cite{hrg}. 
The experimental values plotted are for central Au+Au collisions for $\sqrt{s_{\mathrm {NN}}}$
= 19.6, 62.4 and 200 GeV. The lattice calculations, which predict a CP around 
$\mu_{\mathrm B}$ $\sim$ 300 MeV, are carried out using two-flavor QCD with number of lattice sites
in imaginary time to be 6 and mass of pion around 230 MeV~\cite{latticesus}.
The ratios of the non-linear susceptibilities at finite $\mu_{\mathrm B}$ are obtained
using Pad\'e approximant resummations of the quark number susceptibility series. 
The freeze-out parameters as a function of $\sqrt{s_{\mathrm {NN}}}$ are from~\cite{cleymans}
and $T_{\mathrm c}$ = 175 MeV. 

From comparisons of the experimental data to the HRG model and the lack of non-monotonic
dependence of $\kappa$$\sigma^2$ on $\sqrt{s_{\mathrm {NN}}}$ studied, one 
concludes that there is no indication from the current measurements at RHIC for a CP
in the region of the phase plane with  $\mu_{\mathrm B}$ $<$ 200 MeV. Although
it must be noted that the errors on the experimental data point at 19.6 GeV is 
quite large due to small event statistics. It is difficult to rule out the existence 
of CP for the entire $\mu_{\mathrm B}$ region below 200 MeV. 
The extent to which these results can do that is guided by the theoretical work.
In addition, the expectation of the extent of the critical region  
in $\mu_{\mathrm B}$ is thought to be about 100 MeV. 
The results discussed here form the baseline for the future CP search program at RHIC~\cite{bm09}.
However the fact that the data shows excellent agreement with HRG and Lattice QCD,
both of which assume thermalization, is another non-trivial indication of attainment of 
thermalization (some other measurements are discussed in next section) 
in heavy-ion collisions. Such a conclusion is drawn for the first
time using fluctuation measurements. 

With the idea that the rise and then fall of observables sensitive to CP as  $\mu_{\rm B}$ increases 
should allow us to ascertain the ($T$,$\mu_{\rm B}$) coordinates of the critical point, the
beam energy scan program at  RHIC has started.  The first phase of the 
experimental program at RHIC is expected to be completed in 2010-2011. 
This phase is expected to cover a $\sqrt{s_{\mathrm {NN}}}$ region of 
39 to 5.5 GeV, which corresponds to a $\mu_{\rm B}$ range of 112 to 550 MeV~\cite{bes}.

\section{Hadronic Phase} 

\begin{figure}
\begin{center}
\includegraphics[scale=0.5]{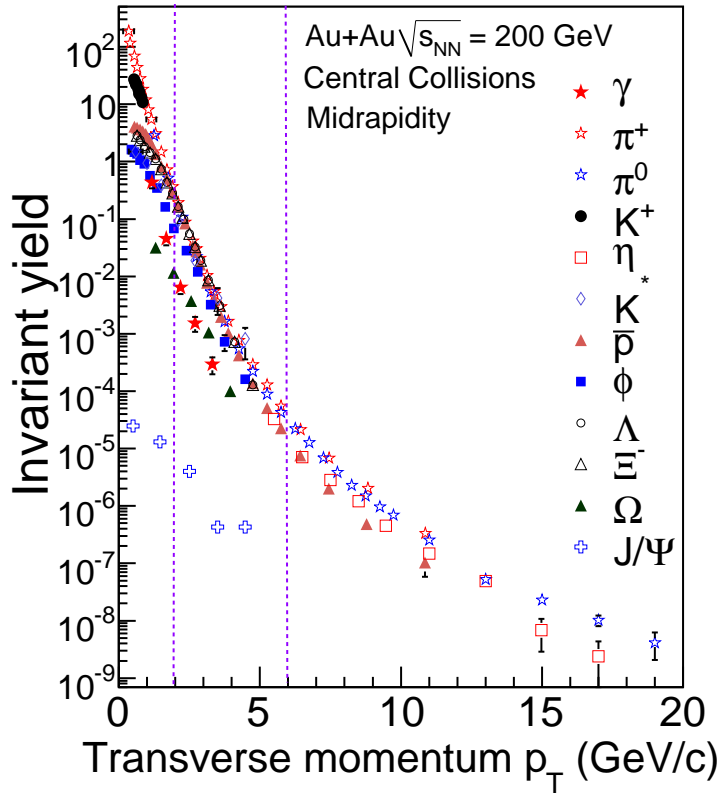}
\caption{(Color online) Compilation of invariant yield of produced particles at midrapidity versus $p_{\mathrm T}$ 
in central Au+Au collisions at $\sqrt{s_{\mathrm {NN}}}$ = 200 GeV ~\cite{starpidprc,phiprl,exptraa,photonraa,rhicspectra}. The lines show the possibility of having
three regions in the spectra where the dominant mechanism of particle productions are different.
See text for more details.}
\label{spectra}
\end{center}
\end{figure}
The measured hadron spectra reflect the properties of the bulk matter at kinetic freeze-out, 
after elastic collisions among the hadrons have ceased. More direct information on the earlier
stages can be deduced from the integrated yields of the different hadron species, which 
change only via inelastic collisions. The point in time at which these inelastic collisions 
cease is referred to as chemical freeze-out, which takes place before kinetic freeze-out. 
RHIC experiments have measured the $p_{\mathrm T}$ distribution of variety of particles over
a wide range of $p_{\mathrm T}$ at midrapidity. A sample of the invariant yield 
($\frac{d^{2}N}{(2\pi p_{\mathrm T})dydp_{\mathrm T}}$ $(GeV/c)^{-2}$) of produced 
particles at RHIC for central Au+Au collisions at 
$\sqrt{s_{\mathrm {NN}}}$ = 200 GeV ~\cite{starpidprc,phiprl,exptraa,photonraa,rhicspectra} is shown in Fig.~\ref{spectra}. Like the 
$p_{\mathrm T}$ dependence of $v_{2}$ discussed in the section 2.3, here also we can separate the spectra into 3 regions based on 
the dominant mechanism of particle production. The low $p_{\mathrm T}$ ($<$ 2 GeV/$c$) is explained 
by thermal model based calculations~\cite{hydroptspectra}, intermediate $p_{\mathrm T}$  (2-6 GeV/$c$) by parton recombination
based approaches~\cite{reco} and high $p_{\mathrm T}$ ($>$ 6 GeV/$c$) by including pQCD based processes or jet production~\cite{vitev_density,xnwang_lifetime}. The only statistical distribution which so far seems to sucessfully describe the $p_{T}$ spectra and $v_2$($p_{\rm T}$) over a wide momentum range is the one based on Tsallis statistics~\cite{xzb}.

In this section we concentrate on low $p_{\mathrm T}$ part of the spectra for rest of the discussions.
The transverse momentum distributions of the different particles contain two components, 
one random and one collective. The random component can be identified as the one that depends 
on the temperature of the system at kinetic freeze-out ($T_{\mathrm {fo}}$). The collective 
component, which arises from the matter density gradient from the center to the boundary of 
the fireball created in high-energy nuclear collisions, is generated by collective flow in the 
transverse direction, and is characterized by its velocity $\beta_T$.
Assuming that the system attains thermal equilibrium, the blast-wave (BW)
formulation~\cite{heniz} can be used to extract $T_{\mathrm {fo}}$ and $\langle \beta_T \rangle$. 
The transverse flow velocity of a particle at a distance $r$ from the center of the emission 
source, as a function of the surface velocity ($\beta_s$) of the expanding cylinder, 
is parameterized as
$\beta_T(r) = \beta_s (r/R)^n$, where $n$ is found by fitting the data. 
The transverse momentum spectrum is then
\begin{eqnarray}
\frac{dN}{p_T \, dp_T} & \propto & \int_0^R r \, dr \, m_T
I_0\left(\frac{p_T \sinh\rho(r)}{T_{\rm{fo}}}\right) \nonumber 
\times K_1\left(\frac{m_T \cosh\rho(r)}{T_{\rm{fo}}}\right),
\label{blasteq}
\end{eqnarray}
where $I_0$ and $K_1$ are modified Bessel functions and $\rho(r) = \tanh^{-1}{\beta_T(r)}$. 

\begin{figure}
\begin{center}
\includegraphics[scale=0.4]{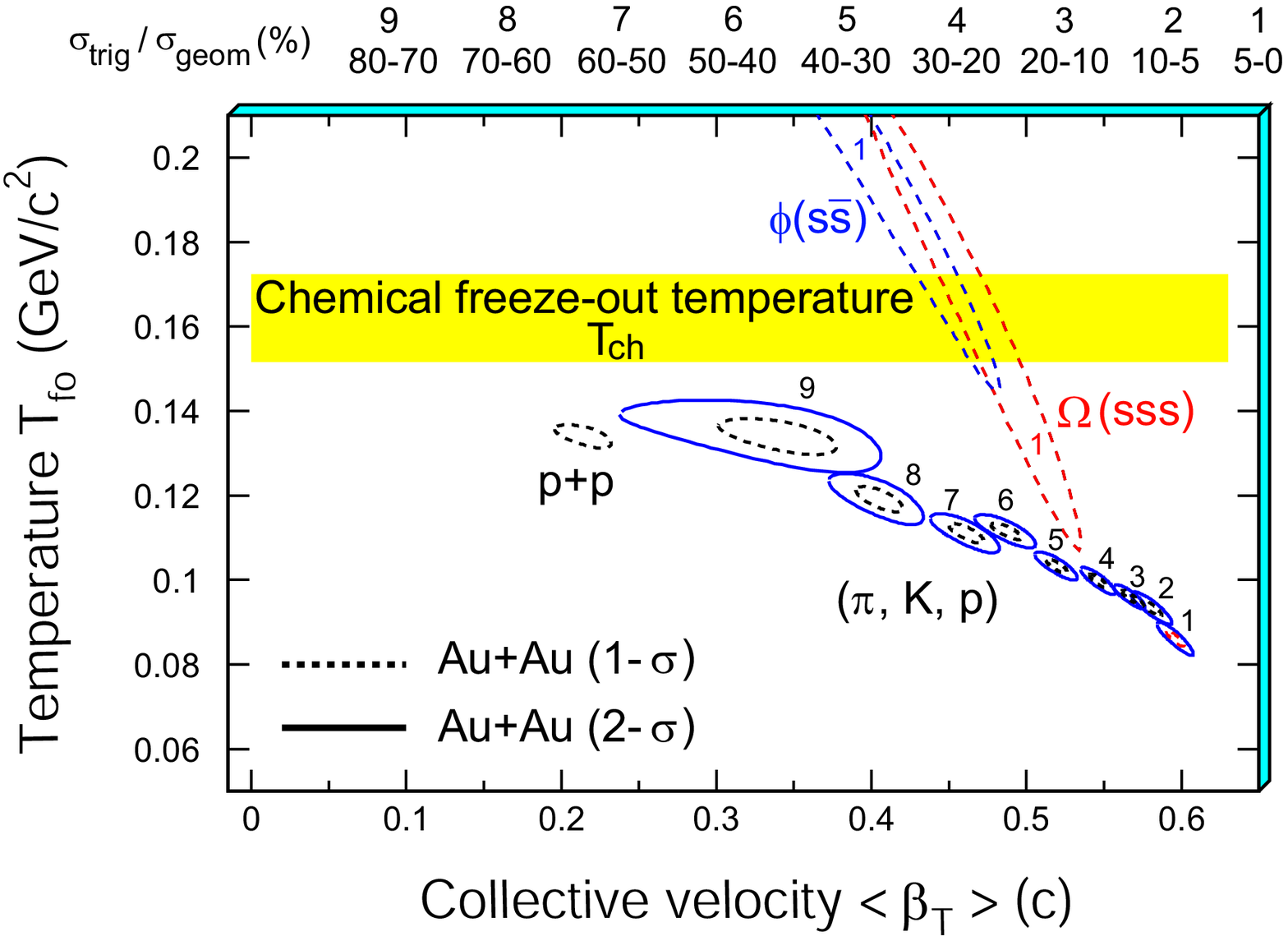}
\includegraphics[scale=0.4]{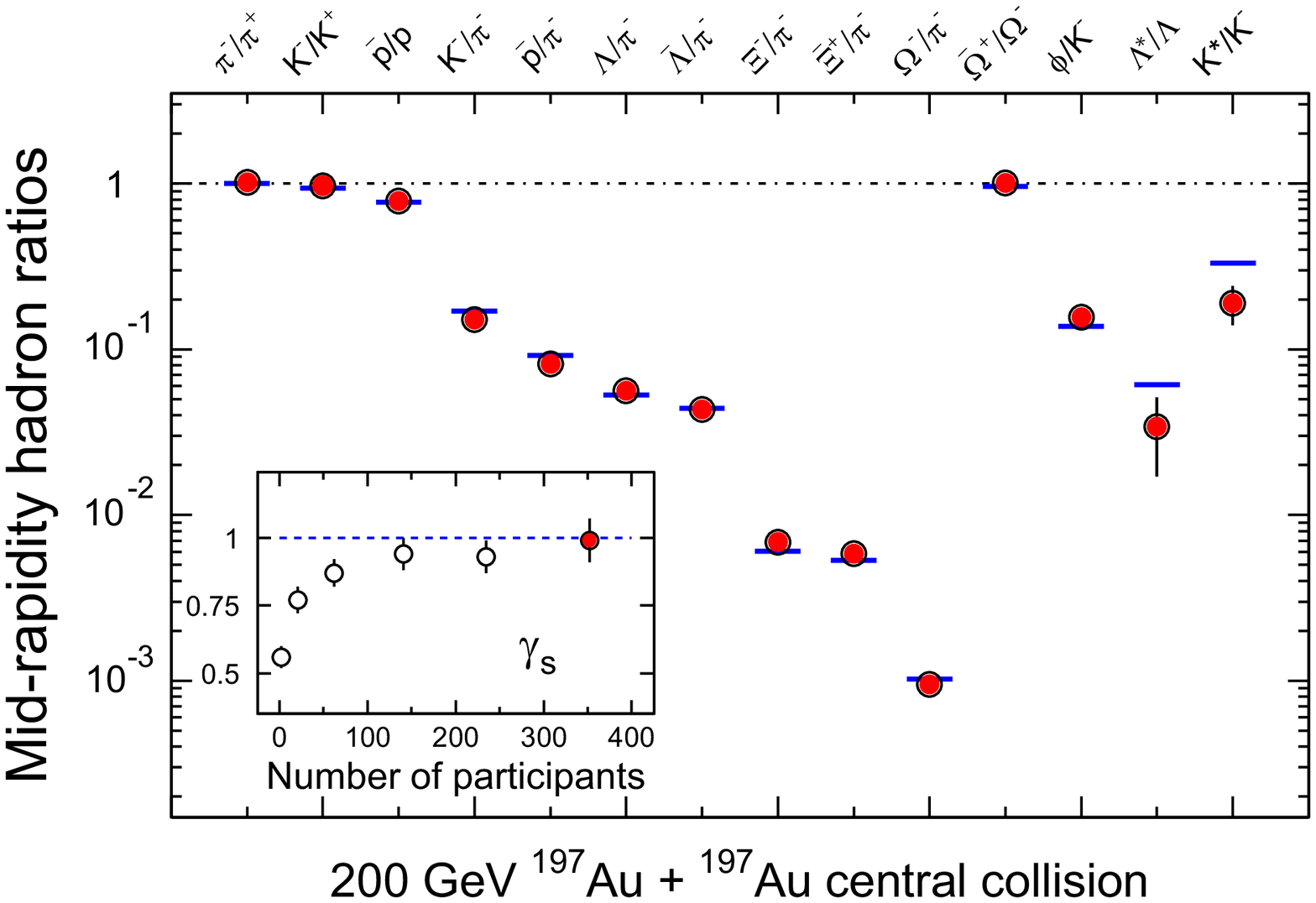}
\caption{(Color online) Left plot: The $\chi^2$ contours, extracted from thermal +
radial flow fits (without allowance for resonance feed-down), for produced
hadrons $\pi, K$ and $p$ and multi-strange hadrons $\phi$ and
$\Omega$. On the top of the plot, the numerical labels indicate
the centrality selection. For $\pi, K$ and $p$, 9 centrality bins
(from top 5\% to 70-80\%) were used for $\sqrt{s_{NN}}=200$ GeV
Au+Au collisions \cite{starpidprc}. The results from p+p
collisions are also shown. For $\phi$ and $\Omega$, only the most
central results are presented. Dashed and
solid lines are the 1-$\sigma$ and 2-$\sigma$ contours,
respectively.
Right plot: Ratios of $p_T$-integrated mid-rapidity yields for
different hadron species measured in STAR for central Au+Au
collisions at $\sqrt{s_{NN}}$ = 200 GeV. The horizontal bars represent
statistical model fits to the measured yield ratios for stable and
long-lived hadrons.  The fit parameters are $T_{ch}=163 \pm 4$
MeV, $\mu_B = 24 \pm 4$ MeV, $\gamma_s = 0.99 \pm 0.07$. The variation
of $\gamma_s$ with centrality is shown in the inset, including the
value (leftmost point) from fits to yield ratios measured by STAR
for 200 GeV p+p collisions.
}
\label{kfo}
\end{center}
\end{figure}

Fits to the identified hadrons $p_{T}$ distributions at midrapidity for Au+Au collisions at  $\sqrt{s_{NN}}$ = 200 GeV
using Eq.~\ref{blasteq} are carried out. The extracted model parameters 
characterizing the random (generally interpreted as a kinetic freeze-out temperature $T_{\rm fo}$) 
and collective (radial flow velocity $\langle \beta_T \rangle$) are shown in  Fig.~\ref{kfo} in
terms of confidence level ($\chi^2$) contours,
for various impact parameters of the collision. 
As the collisions become more and more central, the bulk of the
system, dominated by the yields of $\pi, K, p$ have lower kinetic freeze-out temperature and 
develops stronger collective flow.  On the other hand, even for the most
central collisions, the spectra for multi-strange particles $\phi$
and $\Omega$ appear to reflect a higher freeze-out temperature.

Within a statistical model in thermodynamical equilibrium, the particle abundance in a 
system of volume $V$ can be given by
\begin{equation}
N_i/V=\frac{g_i}{(2\pi)^3}\gamma_{S}^{S_i}\int\frac{1}{\exp\left(\frac{E_i-\mu_BB_i-\mu_SS_i}{T_{\rm {ch}}}\
\right)\pm 1}d^3p\,,
\label{eq:chemical}
\end{equation}
where $N_i$ is the abundance of particle species $i$, $g_i$ is the spin degeneracy, $B_i$ and $S_i$
are the baryon number and strangeness number, respectively, $E_i$ is the particle energy, and the 
integral is taken over all momentum space~\cite{starpidprc}. The model parameters are the chemical 
freeze-out 
temperature ($T_{\mathrm {ch}}$), 
the baryon ($\mu_{\mathrm {B}}$) and strangeness ($\mu_{S}$) chemical potentials, 
and the
{\it ad hoc}
strangeness suppression factor ($\gamma_{S}$).
Measured particle ratios are used to constrain the values of $T_{\mathrm {ch}}$ and $\mu_{\mathrm B}$ 
at chemical freeze-out.

Figure~\ref{kfo} compares STAR measurements of integrated hadron
yield ratios for central Au+Au collisions to statistical model calculations.
The excellent agreement between data and model are observed.
The ratios which include stable and long-lived hadrons through multi-strange baryons, 
is consistent with the light flavors, $u, d,$ and $s$, having reached chemical equilibrium 
(for central and near-central collisions only) at $T_{ch} = 163 \pm 5$ MeV. The 
deviations of the short-lived resonance yields, such as those for $\Lambda ^*$ and K$^*$
collected near the right side of Fig.~\ref{kfo}, from the statistical
model fits, presumably result from hadronic re-scattering after the chemical freeze-out
and needs to be further understood.

The saturation of the strange sector
yields, attained for the first time in near-central RHIC
collisions, is particularly significant. The saturation is
indicated quantitatively by the value obtained for the
non-equilibrium parameter $\gamma_s$ for the strange sector
for central collisions. The temperature deduced from the fits
is essentially equal to the critical value for a QGP-to-hadron-gas
transition predicted by Lattice QCD \cite{lattice}, but is
also close to the Hagedorn limit for a hadron resonance gas,
predicted without any consideration of quark and gluon degrees of
freedom~\cite{hagedorn}. If thermalization is indeed achieved by the bulk matter prior
to chemical freeze-out, then the deduced value of $T_{ch}$
represents a lower limit on that thermalization temperature.

\section{Summary and Outlook} 

\begin{figure}
\begin{center}
\includegraphics[scale=0.5]{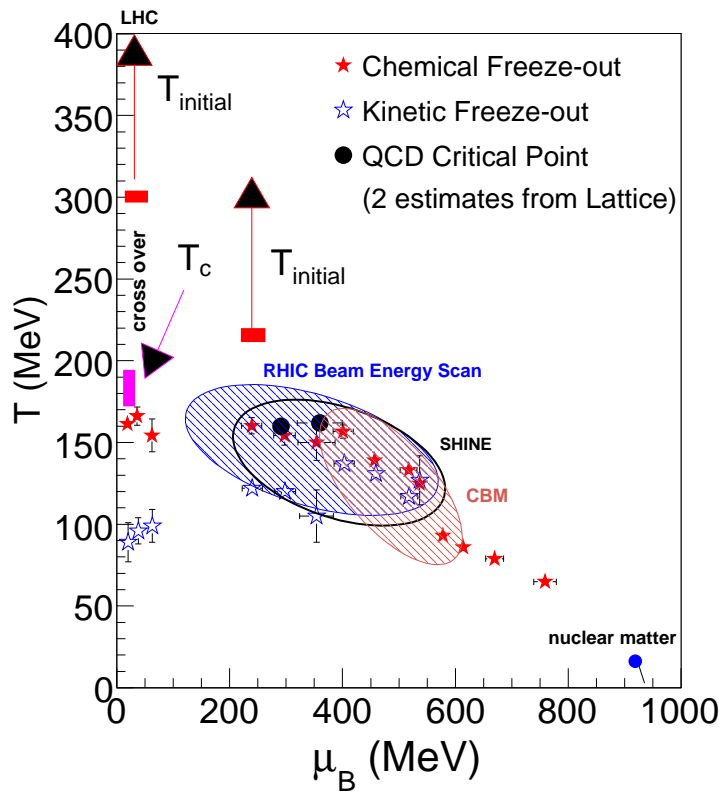}
\caption{(Color online) Temperature vs. baryon chemical potential ($\mu_{\mathrm {B}}$) from heavy-ion
collisions at various $\sqrt{s_{NN}}$~\cite{bm09}. 
The $\mu_{\mathrm {B}}$ values shown are estimated at chemical freeze-out.
The kinetic and chemical freeze-out parameters, extracted using models assuming 
thermal and chemical equilibrium from midrapidity particle ratio and 
$p_{\rm T}$ spectra measurements in heavy-ion collisions.
The range of critical temperatures ($T_{\mathrm {c}}$) of the 
cross-over quark-hadron phase transition at $\mu_{\mathrm {B}}$ = 0
and the QCD critical point from two different calculations from lattice QCD are also indicated. 
Model-based estimates of the range of initial temperature ($T_{\mathrm {initial}}$) 
achieved in heavy-ion collisions based in part on direct photon data at top RHIC~\cite{phenixph} and
SPS~\cite{wa98ph} energies are also shown. 
The range of $\mu_{\mathrm {B}}$ to be scanned in the RHIC beam energy scan 
program corresponding to $\sqrt{s_{NN}}$ = 5.5 to 39 GeV as well as experiments at SPS and CBM
are indicated as shaded ellipse.
The solid point around $T$ $\sim$ 0 and $\mu_{\mathrm B}$ = 938 MeV represents nuclear matter in the ground state.
}
\label{phase}
\end{center}
\end{figure}

In summary, the current understanding of the QCD phase diagram 
is depicted in Fig.~\ref{phase}. From the QCD calculations on lattice it is now 
established theoretically that the quark-hadron transition at $\mu_{B}$ = 0
is a cross-over. The critical temperature for a quark-hadron 
phase transition lies within a range of 170-190 MeV.
Most calculations on lattice also indicate the existence of QCD critical point 
for $\mu_{B}$ $>$ 160 MeV. The exact location of CP is not yet known
unambiguously. Two such predictions computed on lattice are shown in Fig.~\ref{phase}
for a $T_{c}$ of 176 MeV~\cite{criticalpointtheory}. 
New distinct signatures have been predicted by QCD model calculations to locate the critical 
point in the phase diagram. The specific suggestion for a CP search is to look
for non-monotonic variation in the products of the higher moments of net-proton 
and net-charge distributions, which are related to susceptibilities, as a 
function of $\sqrt{s_{NN}}$ (or $T$,$\mu_{B}$). At top RHIC energies, fluctuations 
in net-proton numbers are found to be consistent with expectations from HRG and 
lattice QCD calculations~\cite{starkurtosisprl}. These measurements established  
the baseline for the CP search in the heavy-ion collision program, put constraints 
on the location of CP in the QCD phase plane and added to the evidences for  
thermalization in central Au+Au collisions at RHIC.

High energy heavy-ion collision experiments have seen
distinct signatures which suggest that the relevant degrees of freedom 
in the initial stages of the collisions at top RHIC energies are quark and gluons~\cite{rhicwhitepapers}.
Three such signatures related to strangeness enhancement, jet quenching and partonic collectivity
are discussed in this paper. The initial temperatures ($T_{\mathrm {initial}}$) achieved at top RHIC and SPS
energies are obtained from models~\cite{photonmod} that explain the direct photon 
measurements from the PHENIX experiment at RHIC~\cite{phenixph} and from the WA98 experiment
at SPS~\cite{wa98ph}. From these models, which assume that thermalization is achieved
in the collisions within a time between 0.1--1.2 fm/$c$, the  
$T_{\mathrm {initial}}$ extracted is greater than 300 MeV at RHIC and 
greater than 200 MeV at SPS. Further, the understanding of suppression in high
$p_{T}$ hadron production in heavy-ion collisions relative to $p$+$p$ collisions at RHIC
requires a medium energy density $>>$ 1 GeV/fm$^3$ (critical energy density from lattice
for a phase transition). This also shows that the medium has a high degree of opacity to 
propagation of color charges. In addition the measurement of elliptic flow and the
observation of number of constituent quark scaling demonstrates 
that substantial collectivity has been developed in the partonic phase. The magnitude of
the flow across several hadronic species and a small value of viscosity to entropy ratio 
extracted from the data supports the idea of formation of a strongly 
coupled system in the heavy-ion collisions. This then also supports the notion 
of creating a liquid with low viscosity in high energy nuclear collisions~\cite{perfectliquid}.

The experiments have also measured the temperature at which 
the inelastic collisions ceases (chemical freeze-out) and elastic collisions ceases (kinetic
freeze-out). These temperatures (as shown in Fig.~\ref{phase}) are extracted from the measured 
particle ratios and transverse momentum distributions using model calculations which assume 
the system is in chemical and thermal equilibrium. 

New experimental programs at RHIC, SPS, FAIR and NICA facilities have been designed to explore 
a large part of the QCD phase digram, covering a $\mu_{\mathrm B}$ range of 20-600 MeV. 
Whereas the experimental program at LHC  (probing the cross over region of $\mu_{B}$ $\sim$ 0 MeV 
of the phase diagram) have started to provide an unique opportunity to understand the properties 
of matter governed by quark-gluon degrees of freedom at unprecedented high initial temperatures 
(higher plasma life time) achieved in the Pb+Pb collisions at 2.76-5.5 TeV~\cite{lhc}. Both 
at LHC and RHIC, one specific observable that has the potential to provide a further understanding
of system formed in heavy-ion collision are the dileptons. Theoretically they are from the 
virtual photons, and are different from real photons in having a mass. The dilepton mass opens 
up a new dimension and can
be used to study time evolution of the system in heavy-ion collisions. For example, recently it
has been discussed that virtual photon (dilepton) interferometry provide access to the development 
of collective flow with time~\cite{payal}. Studying the $p_{\rm T}$ dependence of the elliptic flow and 
nuclear modification factor for dileptons for masses corresponding to various hadrons and
beyond will help us understand partonic collectivity and medium opacity. Comparing the spectral
functions of resonances decaying to dileptons and hadrons will let us know about the medium
effects. While the slope of the dilepton $p_{\rm T}$ distributions will tell us about
development of radial flow and provide direct evidence of thermal radiation of partonic
origin in high energy nuclear collisions~\cite{na60}.

\subsection{Acknowledgments}
I would like to thank Drs. J. Alam, S. Chatopadhyay, S. Gupta, V. Koch, L. Kumar, T. K. Nayak, P. K. Netrakanti, D. P. Mahapatra, K. Rajagopal, H. G. Ritter, M. Stephanov, Y. P. Viyogi, N. Xu and Z. Xu for useful discussions. I would like to thank C. Jena, X. F. Luo, Md. Nasim, S. Singha and Drs. M. Sharma, S. Shi and A. Tang for their help in preparation of this manuscript. This work is supported by DAE-BRNS project Scantion No. 2010/21/15-BRNS/2026.


\end{document}